# Polarization analysis of first- and second-order Raman scattering from $MoTe_2$ single crystal.

S. Caramazza,[a] A. Collina,[a] E. Stellino,[a] P. Dore,[b] and P. Postorino[c]

[a] Physics Department, Sapienza University of Rome, P.le Aldo Moro, 00185 Roma, Italy

[b] Physics Department, Sapienza University of Rome, and CNR-SPIN, P.le Aldo Moro, 00185 Roma, Italy

[c] Physics Department, Sapienza University of Rome, and CNR-IOM, P.le Aldo Moro 5, 00185 Roma, Italy

**Abstract**

We report on Raman experiments performed on a single crystal $MoTe_2$ sample. The system belongs to the wide family of Transition Metal Dichalcogenides which includes several of the most interesting two dimensional materials for both basic and applied physics. Measurements were performed in the standard *basal plane* configuration, by placing the *ab* plane of the crystal perpendicular to the wave vector $\mathbf{k_i}$ of the incident beam to explore the in plane vibrational modes, and in the *edge plane* configuration with $\mathbf{k_i}$ perpendicular to the crystal *c* axis, thus mainly exciting out-of-plane modes. For both configurations we performed a polarization-dependent Raman study and we were able to provide a complete assignment of the observed first- and second-order Raman peaks fully exploiting the polarization selection rules. Present findings are in complete agreement with previous first-order Raman data whereas a thorough analysis of the second-order Raman bands, either in *basal-* or *edge-plane* configurations, provides new information and a precise assignment of these spectral structures. In particular, we have observed Raman active modes of the *M* point of the Brillouin zone previously predicted by *ab-initio* calculations and ascribed to either combination or overtone but never previously measured.

## Introduction

Transition Metal Dichalcogenides (TMDs) are layered materials displaying many appealing physical and chemical properties.[1] In particular, they have been used in many applications and devices such as solid lubricants, electrodes in high efficiency photo-electrochemical cells,[2] battery,[3] etc. Among TMDs, Molybdenum based MoX$_2$ (X=S,Se,Te) are semiconductors with indirect energy band-gap in the range of 1.0 -1.3 eV.[1,4,5] They are mostly found in a structure belonging to the symmetry group P6$_3$/*mmc*, with two layers per unit cell. Each layer is composed by a hexagonal plane of Mo atoms sandwiched between two hexagonal planes of chalcogen X atoms, with trigonal prismatic coordination (2H crystal structure).[1] The intra-layer bonding are strong and predominantly covalent, whereas the inter-layer bonding are weak and mainly due to Van der Waals forces.[4,6] Crystal samples can be simply exfoliated and important changes have been observed in the electronic band structure (e.g. crossover from an indirect-to-direct band-gap, gap value) when the crystal thickness is reduced to few layers or down to a single layer.[6-9]. These effects clearly show that the interlayer coupling plays a key role in determining the system properties.[10]

Raman spectroscopy has been widely employed in TMD studies since the position of Raman phonon lines can be used to monitor the number of layers in thin crystal samples,[11] but, in general, it is a powerful tool to get information on both structural and electronic properties of the system.[12-14] We stress that the knowledge of the complete Raman spectrum of a bulk MoX$_2$ sample, including the second-order Raman response, can be of particular importance in achieving information about interlayer couplings, in particular if supported by *ab-initio* calculations.[15,16]

Generally, Raman experiments on MoX$_2$ crystals have been carried out in backscattering geometry with the wave vector ***k**$_i$* of the incident beam perpendicular to the *ab* plane (i.e. parallel to the crystal *c* axis), thus mainly exciting in-plane vibrational modes. Hereafter, this standard scattering configuration will be denoted as *basal plane*. We remark that in this configuration some first- and second-order Raman lines are not allowed by Raman selection rules.[11,15]

In the present work, we focused our attention on bulk 2H-MoTe$_2$, which exhibits an indirect energy band gap of about 1.1 eV[5] and it is emerging as the ideal candidate to be used in valleytronic devices since it shows a strong spin-orbit coupling.[17,18,19] The first-order Raman spectrum of MoTe$_2$ has been investigated in a number of studies, mainly in the case of few-layers samples.[16,20-23] To the best of our knowledge all the available Raman experiments on 2H-MoTe$_2$ have been performed in backscattering geometry in the *basal plane* configuration, while in only one experiment[20] the *right-angle* geometry was employed. Several second-order Raman lines have been observed[16,20,21] and recently assigned on the basis of *ab initio* density functional theory calculations and some other have been only predicted[16] but not experimentally observed. Among these overtones and combination bands involving acoustic vibrational modes have been theoretically discussed.[16] To this respect, it is worth to notice, that the relevance of acoustic vibrational modes in the interpretation of the EXAFS measurements has been recently proved.[24]

Here we collected Raman spectra of $MoTe_2$ single crystals in the standard *basal plane* configuration, but also with the *c* axis perpendicular to the wave vector of the incident beam $\boldsymbol{k_i}$, hereafter mentioned as *edge plane* configuration, thus mainly exciting out-of-plane vibrational modes. To the best of our knowledge, among the number of Raman papers on $MoX_2$, only two works on $MoS_2$ have exploited the *edge plane* configuration.[25,26] It is worth to notice that these two configurations are somehow complementary since in the *edge plane* is possible to see Raman active peaks forbidden in the *basal plane*.

For both *basal* and *edge plane* configurations we performed a polarization-dependent Raman study and evaluated the corresponding Raman selection rules, which allow a detailed phonon assignment. We also carefully measured the second-order Raman spectra collected in both configurations. The experimental results obtained have been discussed in the light of recent theoretical findings,[16] achieving new information on identification and assignment of second-order Raman components.

**Experimental**

Samples used are commercial $MoTe_2$ single crystals (HQ Graphene Inc). We performed micro-Raman measurements at ambient conditions in backscattering geometry with the micro-spectrometer Horiba LabRAM HR Evolution800 equipped with a He-Ne laser (wavelength $\lambda_L$ = 632.8 nm).[27] We notice that the excitation energy in our experiment (1.96 eV) can be considered quasi-resonant with an optical absorption occurring at the M points (see Ref. [16]) and therefore the remarkable signal enhancement allowed us to carefully measure the low intensity second-order Raman signal.
By using a VGB (Volume Bragg Grating) optical filter,[28] we removed the elastic light scattering and collected spectra down to a frequency very close to the laser frequency. Raman spectra in the 20-300 $cm^{-1}$ frequency range were collected by using a Peltier-cooled charge-coupled device (CCD) detector with a spectral resolution better than 1 $cm^{-1}$ thanks to a 1800 grooves/mm grating with 800-mm focal length. For the present measurements, we used a 100x objective with numerical aperture 0.8, and neutral density filters to properly control the laser intensity at the sample surface. We thus obtained a laser spot size on the sample close to 1 μm with a power density of about 0.70

mW/μm$^2$. We verified that under these experimental conditions no appreciable sample heating is observed, consistently with previous literature results.[21,22]

## Results and discussion

The bulk MoTe$_2$ crystal belongs to the space group P6$_3$/mmc (No. 194) and point group D$^4_{6h}$.[1,29] According to the crystal symmetry,[25] the group theoretical analysis shows that the optical phonon modes at the Γ point are twelve: $A_{1g} + 2A_{2u} + 2B_{2g} + B_{1u} + E_{1g} + 2E_{1u} + 2E_{2g} + E_{2u}$, where four phonons ($A_{1g}$, $E_{1g}$ and $2E_{2g}$) are Raman active, one $E_{1u}$ and $A_{2u}$ are acoustic, the other $E_{1u}$ and $A_{2u}$ are infrared active, and $B_{2g}$, $B_{1u}$ and $E_{2u}$ are silent. The frequencies of the Raman active phonons at ambient conditions as reported in different papers[20-22] are given in Table 1. These, as well as present results given for comparison in the last column, will be discussed in the following. The schematic representation of the four Raman active modes and of the $B^1_{2g}$ silent mode is shown in Fig. 1.

In Raman experiments the scattering intensity $I_S$ is given by the relation:[30]

$$I_S \propto |\varepsilon_i\, R\, \varepsilon_s|^2 \qquad (1)$$

where $R$ is the Raman tensor depending on the crystal symmetry and on the particular vibrational mode, and $\varepsilon_i$ and $\varepsilon_s$ are the polarization vectors of the incoming and scattered light, respectively. Knowing the scattering geometry and the properties of the Raman tensor, Eqn (1) allows an *a priori* determination of the selection rules. Hereafter, we will represent the scattering configuration with the sequence $k_i(\varepsilon_i\varepsilon_s)k_s$ in the so-called *Porto notation*:[31] outside the bracket, the symbols show the direction of incident ($k_i$) and scattered ($k_s$) light; inside the brackets, they give the polarization direction of the incident ($\varepsilon_i$) and scattered ($\varepsilon_s$) light. Since our Raman experiments are performed in backscattering geometry, we have: $k_i = -k_s$.

In the present work, the polarization-dependent Raman measurements were realized by keeping fixed both the crystal and the polarization direction of the scattered light $\varepsilon_s$, and by rotating the polarization of the incoming light, that is, by varying the angle θ between $\varepsilon_s$ and $\varepsilon_i$. Specifically, by considering a coordinate system (x y z) such that the crystallographic axes *a* and *b* lie in the xy plane and the *c* axis is along the z axis, $\varepsilon_s$ was kept fixed along the direction of maximum efficiency of the grating and defined as $\varepsilon_s = (1\ 0\ 0)$. The polarization vector $\varepsilon_i$ of the beam impinging on the sample was free to rotate thanks to a λ/2 wave-plate placed along the laser optical path, thus providing $\boldsymbol{\varepsilon_i} = (\cos\theta\ \sin\theta\ 0)$. An analyzer was placed before the grating in order to select the component of the scattered light along the (1 0 0) direction. With this experimental setup, the scattering configurations –z(xx)z (i.e., $\theta = 0$) and –z(yx)z (i.e., $\theta = \pi/2$) correspond to the so-called *polarized* and *depolarized* configurations, respectively. Further details on Raman tensors and selection rules are given in the Supporting Information.

**First order Raman spectra - *Basal plane***

We carried out a series of polarization-dependent measurements on the bulk MoTe$_2$ sample in the *basal plane* configuration ($\boldsymbol{k_i}$ perpendicular to the *ab* crystal plane). Fig. 2 shows a low-magnification photograph of the investigated zone of the sample obtained through the optical microscope. The obtained Raman spectra for the –z(xx)z (*polarized*) and –z(yx)z (*depolarized*) are displayed in Fig. 3a. According to the literature,[16,20-22] the Raman lines clearly detectable in the –z(xx)z polarized spectrum are the $E_{2g}^2$, $A_{1g}$ and $E_{2g}^1$ phonons. The $A_{1g}$ is hardly detectable in the –z(yx)z (*depolarized*) spectrum, while the $E_{1g}$ mode is never detected. Stars mark low intensity spectral features, ascribed to second order Raman components,[16,20] to be analyzed and discussed in the last section. We analyzed the first-order peaks through a standard fitting procedure.[32] We employed Lorentzian functions for $A_{1g}$ and $E_{2g}^1$ Raman lines, a Gaussian function for the low frequency $E_{2g}^2$ peak since its linewidth is very close to the spectral resolution of the spectrometer. The peak frequencies so obtained ($E_{2g}^2$ at 26.9 cm$^{-1}$, $A_{1g}$ at 173.3 cm$^{-1}$, $E_{2g}^1$ at 234.0 cm$^{-1}$) are in good agreement with the literature data, as shown in Table 1.

We remark that a Raman peak associated to the $E_{1g}$ mode cannot be observed for any backscattering geometry from the *basal* plane. However, the observed $E_{1g}$ peak in the experiments in Ref.[21] and [22] (carried out in backscattering) was attributed to a crystal symmetry breaking in three layers MoTe$_2$ samples,[21] and to a faint spectral feature possibly activated by resonance effect (in that case $\lambda_L$ = 532 nm).[22] Finally, the observation of the $E_{1g}$ line was possible in the experiment of Ref.[20] thanks to the use of a *right-angle* instead than a backscattering geometry.

In order to get a deeper understanding of the measured spectra, it is important to notice the strong dependence on polarization of the $A_{1g}$ line, which is very intense in the z(xx)z configuration and hardly detectable in the –z(yx)z configuration. On the contrary, the modes with *E* symmetry give lines with nearly the same intensity (integrated peak area) for both the configurations, as shown in Fig. 3a. This result can be understood by calculating the polarization dependence of the intensity *I* of the $A_{1g}$ and $E_{2g}$ Raman components. By using Eqn (1), one obtains the theoretical predictions for the Raman selection rules (see the Supporting Information for further details):

$$I(A_{1g}) \propto |a\cos\theta|^2 , \qquad I(E_{2g}) \propto |d|^2 , \qquad I(E_{1g}) = 0 \qquad (2)$$

where $a$ and $d$ are constants, $\theta$ is the angle between $\varepsilon_s$ and $\varepsilon_i$ (see above).

From the experimental point of view, by fixing both the position of the crystal and the polarization direction $\varepsilon_s$, we collected Raman spectra by varying $\theta$. Preliminary measurements allows us to set $\theta = 0$ when the maximum intensity of the $A_{1g}$ line is achieved according to Eqn (2). For each $\theta$ value, the intensity of the three Raman peaks was obtained by using the fitting procedure above introduced. The resulting polar plot shown in Fig. 3b results to be in excellent agreement with the best fit curves given by Eqn (2). It is evident that the lines with $E_{2g}$ symmetry do not exhibit any angular dependence, whereas the intensity of the $A_{1g}$ line vanishes in the *depolarized* configuration as expected.

Finally, we focus our attention on the weak peak at 290 cm$^{-1}$ observed in the $-z(xx)z$ configuration, shown in detail the inset of Fig. 3a. As mentioned in different papers,[21,22] this peak can be ascribed to the silent mode $B_{2g}^1$ (see the schematic in Fig. 1) which becomes Raman active in few-layers MoTe$_2$. Its presence has been attributed either to the breaking of translational symmetry along the direction of the $c$ axis[21] or to resonance effects arising from the excitation line ($\lambda_L$ = 633 nm).[22] Notice that the $B_{2g}^1$ mode does not appear in the $-z(xy)z$ configuration, as observed also by Froehlicher et al [22]. In the present case the low intensity $B_{2g}^1$ peak can be observed thanks to the resonance enhancement[16] previously mentioned.

These findings are in remarkable agreement with both previous experimental works reported in literature for other TMDs with 2H structure[11] and the theoretical Raman selection rules we evaluated. This clearly indicates, on one hand, the high quality of the MoTe$_2$ crystal here examined and, on the other hand, the correctness of the experimental and analysis procedures introduced in the present work.

**First order Raman spectra - *Edge plane***

We carried out a polarization dependent Raman study in the *edge plane* configuration, i.e., with the $c$ axis of the crystal perpendicular to the wave vector of the incident beam $k_i$. Fig. 4 shows a photograph of the investigated zone of the sample. In this scattering configuration, since the $c$ axes is no longer parallel to the incident wave vector, the $E_{1g}$ mode is expected to become observable.[15,25,26] In Fig. 5 we compare the Raman spectrum acquired in the *basal plane* with that in the *edge plane* configuration. Stars mark second order Raman components, to be analyzed and discussed in the last section. Both spectra are collected in the scattering condition in which the intensity of the $A_{1g}$ peak is maximum, that is in the *polarized* case for both the *basal* and *edge plane* configurations. From Fig. 5 it is evident the presence of the $E_{1g}$ phonon peaked at 119.6 cm$^{-1}$, in agreement with previous literature results (see Table 1), and that the $A_{1g}$ peak is much more intense than in the *basal plane* case. This effect can be qualitatively attributed to the fact that in the *edge plane* configuration the mostly excited modes are those out-of-plane, where atoms are vibrating along the *c* axes. Hence, an out-of-plane mode like $A_{1g}$ can give a Raman signal much more intense than those given by the in-plane $E_{1g}$ and $E_{2g}$ modes (see Fig. 1).

In the *edge plane* configuration, to define the scattering conditions, we can imagine to rotate the sample around one of its two (x or y) axis in the crystalline plane. For this purpose, starting from the basal plane configuration, we assume the y axis as fixed and we impose a 90° rotation around this axis. We call (x' y' z') the new crystallographic axes (with y'=y). Hereafter, we use this new coordinate system in the Porto notation for the *edge plane* configuration. The -x'(z'z')x' and x'(y'z')x' configurations correspond to *polarized* and *unpolarized* spectra, which are shown in Fig.6a. The corresponding orientation of the crystallographic axes reported in the inset of the figure shows that $k_i$ // x' and that $\varepsilon_i$ and $\varepsilon_s$ lie in the y'z' plane.

In analogy with the procedure adopted for the *basal plane* configuration, keeping fixed both the crystal and the direction of $\varepsilon_s$, we carried out polar-Raman measurements by varying the angle $\theta$ between $\varepsilon_s$ and $\varepsilon_i$. In particular, we set $\theta=0°$ when the polarization vector $\varepsilon_i$ is perpendicular to the *ab* crystallographic plane (i.e., $\varepsilon_i$ // c). The peaks corresponding to the four Raman active modes ($A_{1g}$, $E_{1g}$, $E^1_{2g}$, $E^2_{2g}$) were analyzed as in the *basal plane* case. The obtained polar plot of the corresponding intensities is reported in Fig. 6b, showing that the $A_{1g}$ peak exhibits remarkable polar dependence; in particular, its intensity nearly vanishes in the -x'(y'z')x' *unpolarized* configuration. This explain why the intensity of the $A_{1g}$ peak is maximum in the *polarized* configuration. On the contrary, the intensity of the $E_{1g}$ and $E^1_{2g}$ peaks, which are slightly dependent from polarization, is very small but never vanishes. This behavior is in contrast with the Raman selection rules evaluated for an ideal *edge plane* configuration, as shown in detail in the Supporting Information (see Eqn (S2)), showing that the intensity of the $E_{2g}$ peaks is always null, while both $A_{1g}$ and $E_{1g}$ are strongly dependent on $\theta$. However, present results can be explained by considering a slight misalignment between the crystal axis and the direction of the incident laser beam, i.e. $k_i$ is not perfectly perpendicular to the *c* crystal axis. Indeed, by assuming a 5° misalignment (5° is the value of angle between $k_i$ and x' [see Fig. S3, Supporting Information]), the polar dependence of the $A_{1g}$ and $E_g$ modes (see Eqn (S3), Supporting Information) results to be:

$$I(A_{1g}) \propto |a' \cos\theta|^2 , \qquad I(E_g) \propto |A' \cos\theta|^2 + |B' \sin\theta|^2 \qquad (3)$$

where the constants $a'$, $A'$ and $B'$ are mode-dependent. From Eqn (3) we see that only the intensity of the $A_{1g}$ mode can vanish, whereas the intensities of components with $E_g$ symmetry are always different from zero. Best fit curves are also shown in Fig. 6b as solid lines, in good agreement with the polarization dependence of the experimental intensities of the vibrational modes.

## Second-order Raman spectra

The low intensity spectral features marked by stars appearing in the spectra reported in Figs. 3a, 5 and 6a can be ascribed to the second-order Raman processes on the basis of several literature results.[20,21,33] The *polarized* Raman spectra for *edge* and *basal plane* configurations are reported in Fig. 7 over the 100-240 cm$^{-1}$ frequency range and with an expanded vertical scale. We notice that three peaks of the six predicted in Ref. [19] has been here observed for the first time (see the cyan dotted lines in Fig. 7). The bands at ~143 cm$^{-1}$, ~183 cm$^{-1}$ and ~202 cm$^{-1}$ (see black triangles in Fig. 7) are detectable in both the configurations. They show largely different absolute intensities but about the same relative intensities. These frequency values are well compatible with experimental and theoretical results reported in Ref.[16], where these components are attributed to processes involving a combination/difference or overtone mode of phonons at the *M* point of the Brillouin zone (see the second column of Table 2; here, LA and TA stand for the longitudinal and transverse acoustic modes, respectively). We are not able to ascribe the weak feature at around 187 cm$^{-1}$, more evident in the *basal plane* configuration, to any second-order process (see the vertical line marked with a question mark in Fig. 7).

Looking at Fig. 7, the Raman spectrum of the *edge plane* configuration reveals the presence of other weak spectral bands at 123 cm$^{-1}$, 165 cm$^{-1}$. First column of Table 2 also report the phonon frequencies of the two Raman active modes, $E_{1g}(M)$ (TO mode, 129.2 cm$^{-1}$) and $A_{1g}(M)$ (LO mode, 161.8 cm$^{-1}$) calculated by Guo *et al.*[16] Taking into account the strong enhancement of the signal of both the $E_{1g}$ and $A_{1g}$ peaks we observed in the *edge plane* configuration and the results obtained in Ref.[16], we can reasonably ascribe the weak spectral bands at 123 cm$^{-1}$ and 165 cm$^{-1}$ to the $E_{1g}(M)$ and to the $A_{1g}(M)$ Raman active modes, respectively.

Finally, let us consider again the band at 202 cm$^{-1}$. In Ref.[16], this band has been experimentally found at 205 cm$^{-1}$ (it is worth noticing that also the spectra measured at 1.96 eV and reported in Ref.[16] show the presence of a broad band around 205 cm$^{-1}$) and has not been unambiguously assigned (see Table 2). A close inspection of the spectra of the *edge plane* spectrum shows that this band is given by two components: an intense peak at 202 cm$^{-1}$ and a weaker peak at 207 cm$^{-1}$. By considering, in particular, that the $E_{1g}$(M) peak is rather intense in the *edge plane* configuration, we attributed the first 202 cm$^{-1}$ peak to the $E_{1g}$(M)+TA(M) , the second 207 cm$^{-1}$ peak to the 2LA(M) overtone (see the last column of Table 2).

## Conclusions

We performed a Raman study of the Transition Metal Dichalcogenide MoTe$_2$ by considering the polarization effects on the spectra measured in backscattering geometry on a single crystal sample. By exploiting the characteristics of the employed micro-Raman apparatus, measurements were performed in the frequency range 20-300 cm$^{-1}$ and with a laser spot size on the sample close to 1 μm which allows to exploit the *edge plane* configuration. In the case of measurements performed in the standard *basal plane* configuration, we observed both the expected dominating first-order phonon peaks, and the low intensity second-order components. We also performed measurements by placing the MoTe$_2$ crystal with the *c* axis perpendicular to the wave vector of the incident beam (*edge plane* configuration). The $E_{1g}$ phonon, not observable in the basal plane configuration, becomes well evident in the *edge plane* case. We performed also a detailed polarization analysis of all the first-order components, and obtained polar plots of their intensities. By considering a non-ideal *edge plane* configuration, in which $k_i$ is not perfectly perpendicular to the *c* crystal axis, we obtained a remarkable agreement among the polar plots obtained from experimental data and those predicted by the theoretical Raman selection rules.

A detailed analysis of the second-order Raman spectra, based on recent theoretical results,[16] provided new information on the identification and assignment of second-order Raman components. In particular, we have observed for the first time Raman active modes of the *M* point of the Brillouin zone previously predicted[16] but never observed, and made new assignments.

**Tables:**

| Mode | Ref.[20] (cm⁻¹) | Ref.[21] (cm⁻¹) | Ref.[22] (cm⁻¹) | Present Exp. (cm⁻¹) |
|---|---|---|---|---|
| $E_{2g}^2$ | 25.4 | - | 26.8 | 26.9 |
| $A_{1g}$ | 171.4 | 174 | 173.6 | 173.3 |
| $E_{1g}$ | 116.8 | ∼120 | 119.9 | 119.6 |
| $E_{2g}^1$ | 234.5 | 235 | 234.2 | 234.0 |

**Table 1.** Peak frequencies of the four Raman active modes from literature and from the present experiment, see text.

| Ref.[16] | | | Present work | |
|---|---|---|---|---|
| Calc. freq. (cm⁻¹) | Assignment | Exp. freq. (cm⁻¹) | Exp. freq. (cm⁻¹) | Assignment |
| 129.2 | $E_{1g}(M)$ | - | **122.7** | |
| 132.8/137.4 | $2TA(M)$ or $E_{2g}^1(M) - LA(M)$ | 140.7 | 143.3 | |
| 161.8 | $A_{1g}(M)$ | - | **165.7** | |
| 178.3 | $E_{2g}^1(M) - TA(M)$ | 183.6 | 183.4 | |
| 197.9 | $E_{1g}(M) + TA(M)$ **or** $2LA(M)$ | - | **201.7** | $E_{1g}(M) + TA(M)$ |
| 202.7 | $E_{1g}(M) + TA(M)$ **or** $2LA(M)$ | 205.3 | 207.4 | $2LA(M)$ |

**Table 2.** Phonon frequencies (experimental, calculated) and corresponding assignments from Ref.[16], and from present. *M* indicates modes at the *M* point, LA (TA) stands for longitudinal (transverse) acoustic mode. The last column shows our new assignments for the last two bands.

# Figures:

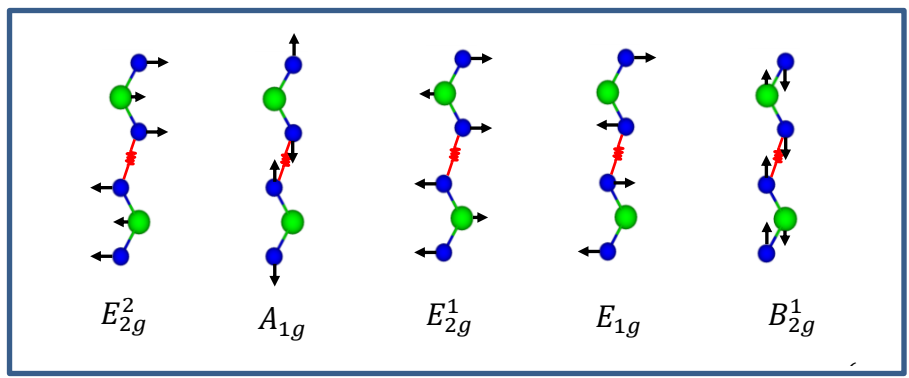

**Fig. 1.** Schematic of the four Raman-active modes and of the silent mode $B_{2g}^1$.

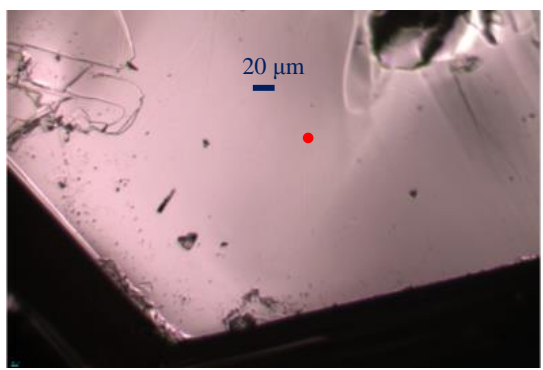

**Fig. 2.** Picture of the MoTe$_2$ sample in the *basal plane* configuration. The red spot is representative of the position where the laser beam impinged the sample.

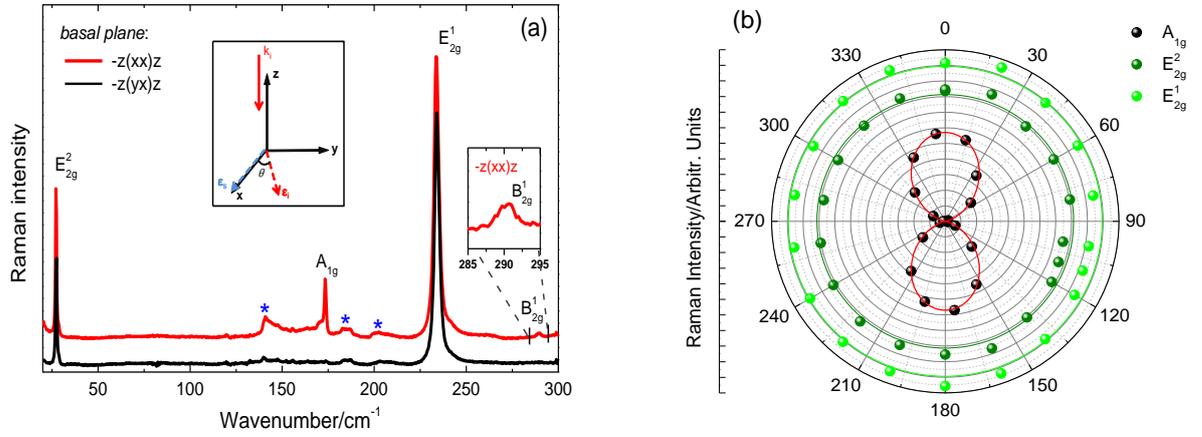

**Fig. 3.** Panel a) *polarized* (-z(xx)z, red) and *depolarized* (-z(yx)z, black) Raman spectra in *basal plane* configuration (crystal axis *c* // $k_i$ ). Spectra are vertically shifted for clarity. Blue stars indicate the second-order Raman components. A schematic of the scattering geometry is shown in the inset. The weak $B^1_{2g}$ peak is shown in the second inset. Panel b): Polar plot of the intensities (integrated peak area) of the $A_{1g}$, $E^1_{2g}$ and $E^2_{2g}$ phonons. By way of comparison, we divided the intensities of the $E^1_{2g}$ mode by a factor 10. Solid lines are the best fit curve of the theoretical selection rules.

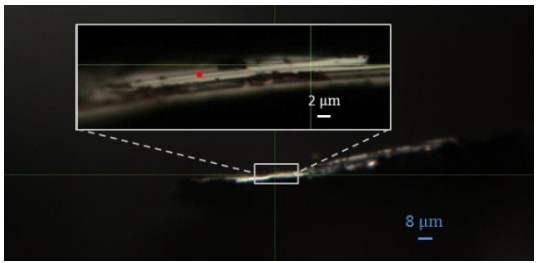

**Fig. 4.** Picture of the MoTe$_2$ sample in the *edge plane* configuration. The red spot is representative of the dimension and of the position where the laser beam impinged the sample.

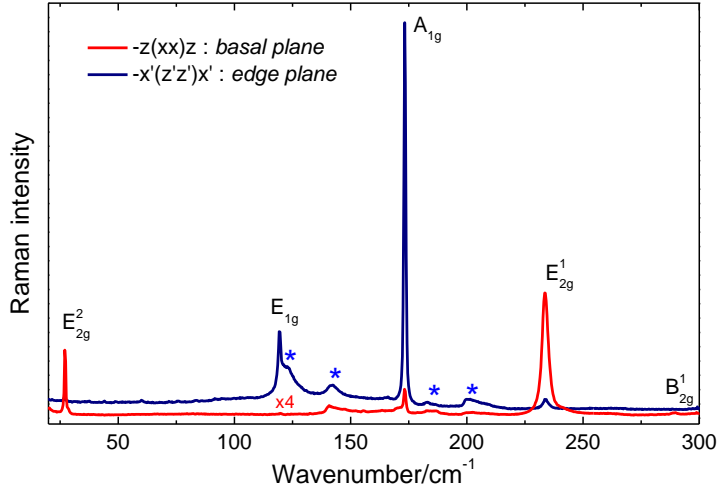

**Fig. 5.** Raman spectra in the two polarized configurations: *basal plane* (–z(xx)z, red) and *edge plane* (–x'(z'z')x', blue). Spectra are vertically shifted and the *basal plane* Raman signal has been multiplied by a factor 4 for clarity. Blue stars indicate the second-order Raman components.

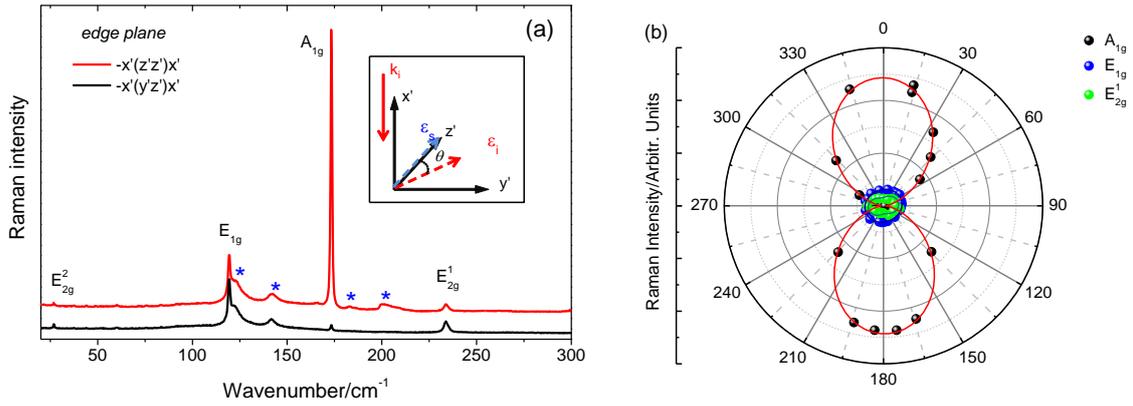

**Fig. 6.** Panel a): *polarized* (–x'(z'z')x', red) and *depolarized* (–x'(y'z')x', black) Raman spectra in *edge plane* configuration. Blue stars indicate second-order Raman components. Spectra are vertically shifted for clarity. A schematic of the new scattering geometry is shown in the inset. Panel b): Polar plot of the intensities (integrated peak area) of the $A_{1g}$, $E_{1g}$ and $E^1_{2g}$ phonons. Solid lines are the best fit curves of the theoretical polarization model. Since the intensities of the $E_{1g}$ and $E^1_{2g}$ Raman components are very weak, corresponding intensity data these modes are crowded close to the origin.

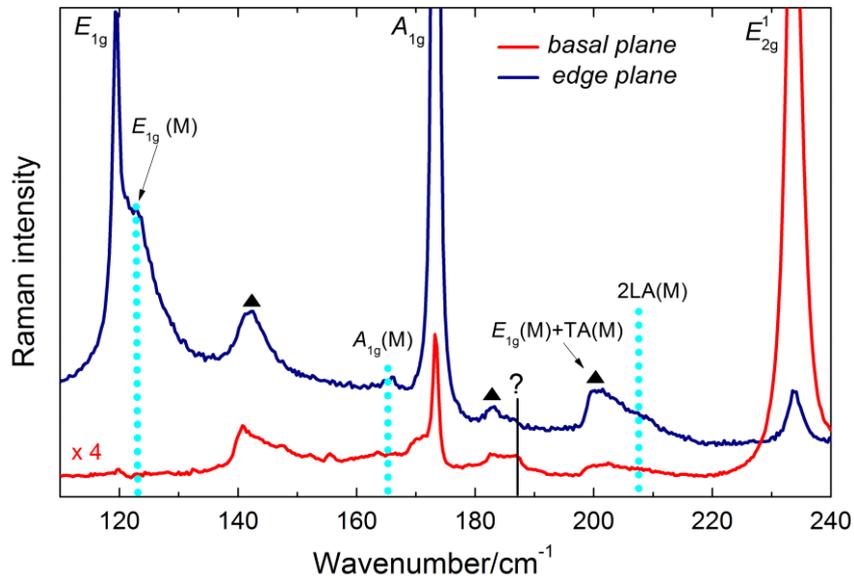

**Fig. 7.** Detail of the polarized Raman spectra in the *edge plane* and *basal plane* configurations. Black triangles indicates the second-order Raman peaks present in both the configuration. LA and TA stand for the longitudinal and transverse acoustic modes, respectively.[16]. Cyan dotted lines mark the three peaks which have been observed in the present experiment only (see Table 2). The peak at 187 cm$^{-1}$ is also indicated.

# Supporting Information

Hereinafter, we denote with (xyz) the coordinated of the laboratory system, whereas the crystallographic axes (in the main text we also referred to the crystallographic axes as: *a b* and *c* axes) are denoted as (x'y'z').

- **Raman tensors in the *basal plane* configuration:**

  In this case the crystallographic axes are coincident with the laboratory system. The Raman tensor in this scattering geometry are:[S1]

$$A_{1g} = \begin{pmatrix} a & 0 & 0 \\ 0 & a & 0 \\ 0 & 0 & b \end{pmatrix}, \quad E_{1g} = \begin{pmatrix} 0 & 0 & 0 \\ 0 & 0 & c \\ 0 & c & 0 \end{pmatrix}, \begin{pmatrix} 0 & 0 & -c \\ 0 & 0 & 0 \\ -c & 0 & 0 \end{pmatrix}; \quad E_{2g} = \begin{pmatrix} d & 0 & 0 \\ 0 & -d & 0 \\ 0 & 0 & 0 \end{pmatrix}, \begin{pmatrix} 0 & -d & 0 \\ -d & 0 & 0 \\ 0 & 0 & 0 \end{pmatrix}. \quad (i)$$

By using $\boldsymbol{\varepsilon_i} \equiv (\cos\theta \ \sin\theta \ 0)$, $\boldsymbol{\varepsilon_s} \equiv (1\ 0\ 0)$ and the Raman tensor (*i*), from Eqn (1) (see main text) we obtained the Eqn (2) reported in the main text:

$I(A_{1g}) \propto |a(\cos(\theta)|^2$

$I(E_{1g}) = 0$ , so the $E_{1g}$ mode is forbidden in this scattering configuration    (S1)

$I(E_{2g}) \propto |d|^2$

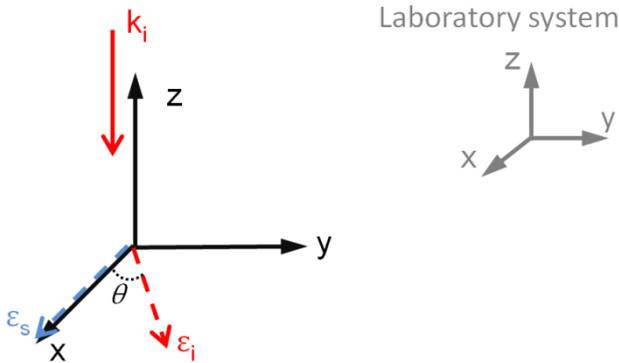

**Fig. S1.** Schematic representation for polarization-dependent measurements in the *basal plane* configurations. (xyz) are the crystallographic axes which coincide with the laboratory system. $k_i$ is the incident wave vector; $\varepsilon_i$ and $\varepsilon_s$ are the polarization vectors of the incoming and scattered light, respectively.

- **Raman tensors in the *edge plane* configuration :**

    In this configuration the new crystallographic axis are denominated as (x'y'z').[1]

    1. If we simulate a rotation of the crystallographic axis of 90° around the y axis we obtained the following Raman tensors[S2] (the orientation of the crystallographic axis in this new geometry is shown in the inset of Fig. 6a, see main text and also in Fig. S2):

$$A_{1g} = \begin{pmatrix} b & 0 & 0 \\ 0 & a & 0 \\ 0 & 0 & a \end{pmatrix}, \quad E_{1g} = \begin{pmatrix} 0 & -c & 0 \\ -c & 0 & 0 \\ 0 & 0 & 0 \end{pmatrix}, \begin{pmatrix} 0 & 0 & c \\ 0 & 0 & 0 \\ c & 0 & 0 \end{pmatrix}; \quad E_{2g} = \begin{pmatrix} 0 & 0 & 0 \\ 0 & -d & 0 \\ 0 & 0 & d \end{pmatrix}, \begin{pmatrix} 0 & 0 & 0 \\ 0 & 0 & -d \\ 0 & -d & 0 \end{pmatrix}.$$

(*ii*)

For the $A_{1g}$, $E_{1g}$ and the $E_{2g}$ we obtain the equations :

$$I(A_{1g}) \propto |b(\cos(\theta)|^2$$
$$I(E_{1g}) \propto |c \sin(\theta)|^2 \qquad \qquad (S2)$$
$$I(E_{2g}) \propto 0$$

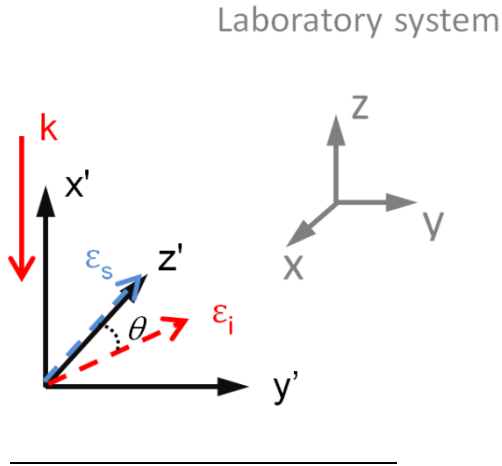

Laboratory system

---

[1] note that in the laboratory system (xyz) we have: $\varepsilon_s \equiv (1\ 0\ 0)$ and $\varepsilon_i \equiv (\cos\theta\ \sin\theta\ 0)$ (the same we used in the case of the *basal plane* configuration). In fact, while the new Raman tensors are express in the (x'y'z') basis, the experimental measurements are referred to the laboratory system (xyz).

**Fig. S2.** Schematic representation for polarization-dependent measurements in the *edge plane* configurations. (x'y'z') are the crystallographic axes; (xyz) is the coordinate system for the laboratory system. $\mathbf{k_i}$ is the incident wave vector; $\boldsymbol{\varepsilon_i}$ and $\boldsymbol{\varepsilon_s}$ are the polarization vectors of the incoming and scattered light, respectively.

2. If we simulate a rotation of the crystallographic axis of 95° around the y axis (instead of 90°, i.e., by assuming a 5° misalignment respect to the ideal *edge plane* configuration), the the Raman tensor (*i*) transform into:[S2]

$$A_{1g} = \begin{pmatrix} a' & 0 & c' \\ 0 & b' & 0 \\ c' & 0 & d' \end{pmatrix}, \quad E_{1g} = \begin{pmatrix} 0 & e' & 0 \\ e' & 0 & f' \\ 0 & f' & 0 \end{pmatrix}, \begin{pmatrix} g' & 0 & h' \\ 0 & 0 & 0 \\ h' & 0 & i' \end{pmatrix}; \quad E_{2g} = \begin{pmatrix} j' & 0 & l' \\ 0 & k' & 0 \\ l' & 0 & m' \end{pmatrix}, \begin{pmatrix} 0 & n' & 0 \\ n' & 0 & o' \\ 0 & o' & 0 \end{pmatrix}.$$

Then, by using Eqn (1), the scattering intensity of the $A_{1g}$, $E_{1g}$, and $E_{2g}$ phonons become:

$$I(A_{1g}) \propto |a'(\cos(\theta)|^2$$

$$I(E_{1g}) \propto |e'(\sin\theta)|^2 + |g'(\cos(\theta)|^2 \quad \text{(S3)}$$

$$I(E_{2g}) \propto |n'(\sin\theta)|^2 + |j'(\cos(\theta)|^2$$

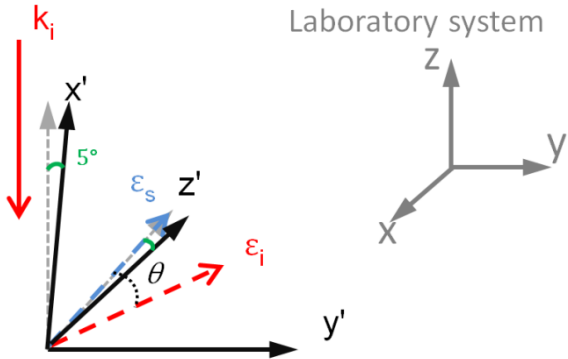

**Fig. S3.** Schematic representation for polarization-dependent measurements in the *edge plane* configurations with a small (5°) misalignment respect to the ideal *edge plan*e configuration. $\mathbf{k_i}$ is the incident wave vector; $\boldsymbol{\varepsilon_i}$ and $\boldsymbol{\varepsilon_s}$ are the polarization vectors of the incoming and scattered light, respectively. 5° is the angle between $\mathbf{k_i}$ and x', which is the same between –x (laboratory system) and z' (crystallopgraphic axis).

------